\documentclass[prl,twocolumn,showpacs,amsfonts,amsmath,floatfix]{revtex4}
\usepackage{graphicx}
\usepackage{psfrag}
\newcommand{\Journal}[4]{#1 {\bf #2}, #3 (#4)}
\newcommand{\PR}{Phys. Rev.}
\newcommand{\PRL}{Phys. Rev. Lett.}
\newcommand{\PRA}{Phys. Rev. A}

\newcommand{\JMP}{J. Math. Phys.}

\newcommand{\Science}{Science}
\newcommand{\PLA}{Phys. Lett. A}
\begin{document}
\title {Soluble Models of Strongly Interacting Two-level Ultracold Gases in 
Tight Waveguides with Coupling to the Quantized Electromagnetic Field}
\author{M. D. Girardeau}
\email{girardeau@optics.arizona.edu}
\affiliation{College of Optical Sciences, University of Arizona, 
Tucson, AZ 85721, USA}
\date{\today}
\begin{abstract}
A generalized Fermi-Bose mapping method is used to determine the exact
ground states of six models of strongly interacting ultracold gases 
of two-level atoms in tight waveguides, which are generalizations
of the Tonks-Girardeau (TG) gas (1D Bose gas with point hard cores) 
and fermionic Tonks-Girardeau (FTG) gas (1D spin-aligned
Fermi gas with infinitely strong zero-range attractions). Three of these
models exhibit a quantum phase transition in the presence of an external
magnetic field, associated with a cooperative ground state rearrangement 
wherein Fermi energy is traded for internal excitation energy. 
After investigation of these models in the absence of an electromagnetic field,
one is generalized to include resonant interactions with a single photon 
mode, leading to a possible thermal phase transition associated with Dicke 
superradiance.   
\end{abstract}
\maketitle
The rapidly increasing sophistication of experimental techniques for probing 
ultracold gases has caused a shift of emphasis in theoretical and experimental 
work in recent years, from effective field approaches to more refined 
methods capable of dealing with strong correlations. In ultracold gases 
confined in de Broglie waveguides with transverse trapping 
so tight that the atomic dynamics is essentially one-dimensional \cite{Ols98}, 
with confinement-induced resonances \cite{Ols98,GraBlu04} 
allowing Feshbach resonance tuning \cite{Rob01} of the effective 1D 
interactions to very large values, such correlations are greatly
enhanced. This has allowed experimental verification 
\cite{Par04Kin04,Kin05,Kin06} of the fermionization of bosonic ultracold 
vapors in such geometries predicted by the Fermi-Bose (FB) 
mapping method \cite{Gir60Gir65}. Here 
the FB mapping method used in \cite{Gir60Gir65} to solve the TG gas and in
\cite{GraBlu04,GirOls03,GirNguOls04} to solve the FTG gas is generalized to 
obtain exact solutions of six models of strongly interacting 1D ultracold
gases of 2-level atoms in tight waveguides, with and without resonant coupling 
to a single mode of the quantized electromagnetic field. 

Consider first some 1D gases of ultracold two-level atoms with no coupling to 
the quantized electromagnetic field. The procedure for 
obtaining the exact solutions is similar to that used in recent work 
\cite{GirMin07}
on two-component mixtures of 1D gases, but with an 
important difference: In a mixture of dissimilar species 
there is no symmetry requirement on exchange of dissimilar
atoms, whereas in the present case of a single species with two
internal levels g (ground) and e (excited), the wave function must be
symmetric (Bose) or antisymmetric (Fermi) under combined exchange of spatial
coordinates and internal states. Starting as in \cite{GirMin07} 
from ``model states'' $\Psi_M$ which are ideal Fermi or Bose gases, one can 
generate the interactions by a mapping $\Psi_M\to\Psi=A\Psi_M$ 
where $A$ is $\pm 1$ everywhere and has discontinuities at pair 
contact which introduce zero-range TG and/or FTG interactions. 

Denote spatial coordinates of the $N$ atoms by $x_1,\cdots,x_N$ and their
internal state labels by $s_1,\cdots,s_N$ where each $s_j$ takes on the
values $g$ or $e$. Models I, II, and III start from the model states
$\Psi_{\alpha M}^F$ of an ideal Fermi gas, and models IV, V, and VI from 
those $\Psi_{\alpha M}^B$ of an ideal Bose gas: 
\begin{equation}\label{Psi_model}
\Psi_{\alpha M}^F=\text{det}_{j,\ell=1}^N\phi_{\nu_j}(x_\ell,s_\ell)
\ ,\ \Psi_{\alpha M}^B=\text{alt}_{j,\ell=1}^N\phi_{\nu_j}(x_\ell,s_\ell)
\end{equation}
where $\text{alt}$ denotes an alternant (minus signs in the determinant
$\text{det}$ replaced by plus signs). 
The orbitals $\phi_{\nu}(x,s)$ are a complete set of energy eigenstates of a 
single-particle Hamiltonian 
$\hat{H}(x,s)=-\frac{\hbar^2}{2m}\frac{\partial^2}{\partial x^2}
+v(x,s)+\delta_{sg}\epsilon_g+\delta_{se}\epsilon_e$  
appropriate to  
given boundary conditions and external potential $v(x,s)$, satisfying the
orthonormality relation 
$\sum_{s=g,e}\int dx\phi_{\nu}^*(x,s)\phi_{\nu'}(x,s)=\delta_{\nu\nu'}$. 
The $\phi_{\nu_j}$ in the Slater determinant $\Psi_M^F$ are a selection of $N$ 
different $\phi_\nu$, whereas those in the alternant $\Psi_M^B$ are such a 
selection but
with multiple occupancy allowed. The 
terms depending on the internal energy levels $\epsilon_g$ and 
$\epsilon_g$ account for the energy
difference $\epsilon_e-\epsilon_g$ between ground and excited states.
$\Psi_M^F$ and $\Psi_M^B$ are eigenstates of the interaction-free model 
Hamiltonian $\hat{H}_M=\sum_{j=1}^N\hat{H}(x_j,s_j)$ with eigenvalue
$\sum_{j=1}^N\epsilon_{\nu_j}$, and all selections of the $\nu_j$ give all
such $N$-atom energy eigenstates. The Bose ground state is a completely
Bose-Einstein condensed state with all $N$ atoms in the lowest orbital
$\nu=0$, i.e., $\Psi_{0M}^B=\prod_{j=1}^N\phi_0(x_j,s_j)$, and the Fermi
ground state is a Slater determinant of the lowest $N$ orbitals, i.e.,
$\Psi_{0M}^F=\text{det}_{\nu,\ell=(0,1)}^{(N-1,N)}\phi_{\nu}(x_\ell,s_\ell)$.
Assume harmonic trapping with the same longitudinal frequency for both
internal ground and excited atoms, i.e., $v(x,s)=\frac{1}{2}m\omega^2x^2$. 
Then if $\epsilon_{eg}=\epsilon_e-\epsilon_g<N\hbar\omega$,
the state with all $N$ atoms in internal level g is not the ground state, 
because the energy can be lowered by exciting the atom at the top of the
Fermi sea to internal level e and moving it to the lowest harmonic oscillator
level, signalling an instability of the putative ground state against 
internally exciting atoms near the top of the Fermi sea and moving  them to 
levels near the bottom, which become doubly occupied with one g-atom and one 
e-atom. If $\epsilon_{eg}$ is a hyperfine splitting, this inequality cannot be 
satisfied for currently achievable values of $N$, so in the absence of an 
external magnetic field the $N$-atom ground states of these models consist of 
atoms in their ground levels g. However, in the
presence of an external magnetic field in the proper direction one has 
$\epsilon_{eg}=\epsilon_{\text{HF}}-\epsilon_{\text{Zeeman}}$ where 
$\epsilon_{\text{HF}}$ and $\epsilon_{\text{Zeeman}}$ are the hyperfine and
Zeeman splittings. This can be made arbitrarily small by tuning the
magnetic field, so a quantum phase transition of this type should be
experimentally realizable in a TG gas. It also has intrinsic theoretical
interest because it is induced by the strong interatomic interactions and
cannot occur in an ideal Fermi gas.

The energy eigenstates including TG and/or 
FTG interatomic interactions are generated by multiplication of the ideal 
Fermi and Bose gas states by appropriate mapping functions
$A(x_1,s_1,\cdots,x_N,s_N)$. The original FB mapping solution
for the TG gas \cite{Gir60Gir65} starts from a model state $\Psi_F$ which 
is an ideal spin-aligned Fermi gas Slater determinant, and generates the
energy eigenstates $\Psi_B$ of a system of bosons with TG (point hard core) 
interactions by multiplying $\Psi_F$ by a ``unit antisymmetric mapping 
function''
\begin{equation}\label{1960map}
A(x_1,\cdots,x_N)=\prod_{1\le j<\ell\le N}\text{sgn}(x_j-x_\ell) 
\end{equation}
where the sign function $\text{sgn}(x)$ is $+1\ (-1)$ if $x>0\ (x<0)$. This 
changes the antisymmetry nodes at $x_j=x_{\ell}$ into collision cusps which are
the $c\to +\infty$ limit of the collision cusps in the Lieb-Liniger solution 
for a Bose gas with interactions $c\delta(x_j-x_{\ell})$ \cite{LieLin63}.
The FTG gas is a ``mirror image'' of the TG gas, consisting of spin-aligned
fermions with infinitely-strong zero-range attractions which are a zero-width,
infinite depth limit of a square well of depth $V_0$ and width $2x_0$, with 
the limit taken such that $V_0 x_0^2\to(\pi\hbar^2)/8\mu$ where $\mu$ is the 
effective mass of the colliding pair \cite{GraBlu04,GirOls03,GirNguOls04}. 
It causes odd-wave scattering (1D analog of 3D p-wave scattering) with 1D 
scattering length $a_{1D}=-\infty$, with the result that all energy eigenstates
$\Psi_F$ of the FTG gas are obtained from corresponding ideal Bose gas states
$\Psi_B$ (the model states) by the mapping $\Psi_B\to\Psi_F=A\Psi_B$, where
$A$ is exactly the same mapping (\ref{1960map}), which 
now introduces sign-changing discontinuities in the FTG states necessary to
reconcile fermionic antisymmetry with a strong interaction
in the zero-range limit $x_0\to 0$ \cite{CheShi98}. Inside the square well the 
solution passes smoothly through a zero at $x_j-x_\ell=0$, so the discontinuity
is an illusion of the zero-range limit 
\cite{CheShi98,GraBlu04,GirOls03,GirNguOls04}. 

{\it Model I:} This is a Bose gas with TG gg and ee interactions and FTG
ge interactions. Its energy eigenstates $\Psi_\alpha^B$ are generated by 
mapping from the two-level ideal Fermi gas model states $\Psi_{\alpha M}^F$ of 
Eqs. (\ref{Psi_model}) according to $\Psi_\alpha^B=A\Psi_{\alpha M}^F$
where the mapping function $A$ is the simple one (\ref{1960map}) of the
original TG gas solution \cite{Gir60Gir65}. This 
generates TG gg and ee interactions, but FTG ge interactions. This is 
easily seen by a simple example. Suppose that g and e atoms see the same
harmonic trap potential $v(x,s)=\frac{1}{2}m\omega^2x^2$. Then the lowest state
$\Psi_{\alpha M}^F$ with $2$ atoms in the ground level g and $1$ atom in the
excited level e is a Slater determinant
(\ref{Psi_model}) constructed from the three harmonic oscillator orbitals 
$\phi_\nu(x,s)$ where $\phi_0(x,s)=u_0(x)\delta_{sg}$, 
$\phi_1(x,s)=u_0(x)H_1(Q)\delta_{sg}$, and $\phi_2(x,s)=u_0(x)\delta_{se}$, 
where $Q=x/x_{\text{osc}}$, $x_{\text{osc}}=\sqrt{\hbar/m\omega}$, 
$u_0(x)=\text{const.}e^{-Q^2/2}$, and $H_n$ are Hermite polynomials, as in 
\cite{GirWriTri01}. Dropping a normalization constant and multiplying by 
$A$ of Eq. (\ref{1960map}) to obtain $\Psi^B$, one finds
\begin{eqnarray}\label{modelI}
\hspace{-0.5cm}&&\Psi^B=u_0(x_1)u_0(x_2)u_0(x_3)\nonumber\\
\hspace{-0.5cm}&&\times[-|x_1-x_2|\text{sgn}(x_1-x_3)\text{sgn}(x_2-x_3)
\delta_{s_1g}\delta_{s_2g}\delta_{s_3e}\nonumber\\
\hspace{-0.5cm}&&+|x_1-x_3|\text{sgn}(x_1-x_2)\text{sgn}(x_2-x_3)
\delta_{s_1g}\delta_{s_3g}\delta_{s_2e}\nonumber\\
\hspace{-0.5cm}&&-|x_2-x_3|\text{sgn}(x_1-x_2)\text{sgn}(x_1-x_3)
\delta_{s_2g}\delta_{s_3g}\delta_{s_1e}]\ .
\end{eqnarray}
If atoms 1 and 2 are both in level g, then $\Psi^B$ has a TG 
$|x_1-x_2|$ cusp at $x_1=x_2$, but 
if atom 1 is in level g and atom 2 in level e or vice
versa, then 
$\Psi^B$ has a FTG discontinuity at $x_1=x_2$. Similar remarks apply to the 
atom pairs (1,3) and (2,3).
This generalizes to eigenstates $\Psi_\alpha^B$ with arbitrary numbers
of atoms in levels g and e, showing that two atoms both in level g or both in
level e have TG cusps at their collision points $x_j=x_\ell$, but if one is
in level g and the other in level e, there is a FTG discontinuity instead.
This model is equivalent to that of Mousavi {\it et al.} \cite{MCLM07},
although the equivalence is not immediately apparent since the notation in
\cite{MCLM07} is very different from that used here, and the interactions
in \cite{MCLM07} are defined differently, as TG in the three triplet gg, ee,
and ge channels and FTG in the singlet ge channel \cite{Note1}. To see the 
equivalence note that (\ref{modelI}) can be 
rewritten in terms of triplet and singlet ge channels as \cite{Note2}
\begin{eqnarray}\label{Mousavi}
\hspace{-0.8cm}&&\Psi^B=\frac{1}{4}u_0(x_1)u_0(x_2)u_0(x_3)
\{|x_1-x_2|\nonumber\\
\hspace{-1cm}&&\times\text{sgn}(x_1-x_3)\text{sgn}(x_2-x_3)
\delta_{s_3g}(\delta_{s_1g}\delta_{s_2e}+\delta_{s_2g}\delta_{s_1e})\nonumber\\
\hspace{-1cm}&&-|x_1-x_3|\text{sgn}(x_1-x_2)\text{sgn}(x_2-x_3)\nonumber\\
\hspace{-1cm}&&\times\delta_{s_2g}
(\delta_{s_1g}\delta_{s_3e}+\delta_{s_3g}\delta_{s_1e})\nonumber\\
\hspace{-1cm}&&+|x_2-x_3|\text{sgn}(x_1-x_2)\text{sgn}(x_1-x_3)\nonumber\\
\hspace{-1cm}&&\times\delta_{s_1g}
(\delta_{s_2g}\delta_{s_3e}+\delta_{s_2g}\delta_{s_3e})\nonumber\\
\hspace{-1cm}&&+\text{sgn}(x_1-x_2)\text{sgn}(x_1-x_3)
\text{sgn}(x_2-x_3)\nonumber\\
\hspace{-1cm}&&\times[(x_1-2x_3+x_2)\delta_{s_3g}
(\delta_{s_1g}\delta_{s_2e}-\delta_{s_2g}\delta_{s_1e})\nonumber\\
\hspace{-1cm}&&+(x_3-2x_2+x_1)\delta_{s_2g}
(\delta_{s_3g}\delta_{s_1e}-\delta_{s_1g}\delta_{s_3e})\nonumber\\
\hspace{-1cm}&&+(x_2-2x_1+x_3)\delta_{s_1g}
(\delta_{s_2g}\delta_{s_3e}-\delta_{s_3g}\delta_{s_2e})]\}\ .
\end{eqnarray}
It follows that the interaction in the
triplet ge channels is of TG form \cite{Note3} whereas that in the singlet
ge channels is of FTG form, but the equal expression (\ref{modelI}) is much 
simpler due to
cancellations in (\ref{Mousavi}), does not require separation into singlet
and triplet channels, yet exhibits the TG interaction in the triplet gg
channel, which is not evident in the form (\ref{Mousavi}). 
The model represents a two-level hybrid TG-FTG 
gas with infinitely strong repulsive gg and ee TG interactions and infinitely
strong \emph{attractive} ge FTG interactions, although it was called simply a
two-level TG gas in \cite{MCLM07}. Its experimental realization would be
difficult since generation of the FTG interaction would require a p-wave ge
resonance, and one would have to simultaneously create TG gg and ee 
interactions, although the latter could perhaps be done by the methods of 
\cite{Par04Kin04,Kin05,Kin06}. 

{\it Model II:} This is a Bose gas with TG gg and ee interactions but no
ge interactions. It starts from the same ideal Fermi gas model states
$\Psi_{\alpha M}^F$ of Eqs. (\ref{Psi_model}) and generates the 
states $\Psi_\alpha^B$ with TG gg and ee interactions by a mapping 
$\Psi_\alpha^B=A\Psi_{\alpha M}^F$, but the mapping is now a more complicated
one depending on both spatial and internal variables, which is everywhere
$\pm 1$ and antisymmetric 
under exchanges $(x_j,s_j)\leftrightarrow (x_\ell,s_\ell)$:
\begin{eqnarray}\label{XTonksmap}
&&\hspace{-0.7cm}A(x_{1},s_{1};\cdots;x_{N},s_{N})=\prod_{1\le j<\ell\le N}
\alpha(x_{j},s_{j};x_{\ell},s_{\ell}) \nonumber\\
&&\hspace{-0.7cm}\alpha(x_{j},s_{j};x_{\ell},s_{\ell})\nonumber\\
&&\hspace{0.7cm}=\delta_{s_{j},s_{\ell}} 
\text{sgn}(x_{j}-x_{\ell})
+\delta_{s_{j}g}\delta_{s_{\ell}e}-\delta_{s_{j}e}\delta_{s_{\ell}g}\ .
\end{eqnarray}
This is the same mapping used previously in a model of a TG Mach-Zehnder
interferometer, where the state variable $s$ distinguished between the
two interferometer arms \cite{GirDasWri02}. It generates
a TG interaction in the gg and ee channels $\delta_{s_{j},s_{\ell}}$,
but no ge interaction. It should be easier to realize 
experimentally than model I.

{\it Model III:} This is a \emph{Fermi} gas with FTG ge interactions but
no gg or ee interactions. It again starts with the ideal Fermi gas model states
$\Psi_{\alpha M}^F$ of Eqs. (\ref{Psi_model}), but uses a different mapping
$A$, which is everywhere $\pm 1$ and is now \emph{symmetric} under exchanges 
$(x_j,s_j)\leftrightarrow (x_\ell,s_\ell)$, to generate the states 
$\Psi_\alpha^F$ with FTG interactions: 
\begin{eqnarray}\label{symmetricmap}
&&\hspace{-0.3cm}A(x_{1},s_{1};\cdots;x_{N},s_{N})=\prod_{1\le j<\ell\le N}
\alpha(x_{j},s_{j};x_{\ell},s_{\ell})\nonumber\\
&&\hspace{-0.3cm}\alpha(x_{j},s_{j};x_{\ell},s_{\ell})
=\delta_{s_j,s_\ell}\nonumber\\ 
&&\hspace{1.5cm}+(\delta_{s_jg}\delta_{s_{\ell}e}
-\delta_{s_je}\delta_{s_{\ell}g})\text{sgn}(x_{j}-x_{\ell})\ .
\end{eqnarray}
It generates no interaction in the gg and ee channels 
$\delta_{s_j,s_{\ell}}$, but FTG ge interactions. 

{\it Models IV, V, and VI:} These all start from the ideal two-level Bose gas 
model states $\Psi_{\alpha M}^B$ of Eqs. (\ref{Psi_model}). Model IV uses the
simple mapping (\ref{1960map}) and has FTG gg, ee, and ge interactions,
model V uses the mapping (\ref{XTonksmap}) and has FTG gg and ee interactions,
and model VI uses the mapping (\ref{symmetricmap}) and has FTG ge interactions.
Model VI should be the easiest to realize since a p-wave Feshbach resonance
is required only in the ge channel.

{\it Interaction with a coherent photon mode:} Some thirty years
ago I suggested the possibility of a cooperative coupling between Bose-Einstein
condensation (BEC) in liquid $^4$He and BEC of van der Waals virtual photons 
into a superradiant mode, and constructed a simple model exhibiting such a 
quantum phase transition \cite{Gir78}. Such equilibrium superradiance has 
never been observed (nor, I believe, searched for) in superfluid $^4$He, but in
recent years coupling of matter waves and a superradiant photon mode have
been observed in ultracold Bose gases \cite{Ino99,Sch03}, and it has been 
pointed out that essentially the same phenomena can also occur in ultracold 
Fermi gases \cite{MooMey01,KettIno01}. In fact, it had been shown long ago
by Hepp and Lieb \cite{HepLie73} that BEC of the atoms is not necessary for
cooperative coupling of internal and translational degrees of freedom
in a multilevel gas via a superradiant photon mode. Motivated by these results,
model I above will be generalized by adding resonant coupling to a single 
mode of the quantized electromagnetic field \cite{Note5}. 

Start with the atom-field interaction Hamiltonian in the electric field gauge 
and rotating wave approximation:
$\hat{H}_{\text{atom-field}}=i\sum_{j=1}^N\sum_{k\lambda}
\sqrt\frac{\hbar ck}{L}(\mathbf{d}_{eg}\cdot\mathbf{e}_{k\lambda}
\hat{S}_j^+\hat{b}_{k\lambda}e^{ikx_j}-\text{H.c.})$.
Here $\hat{b}_{k\lambda}$ and $\hat{b}_{k\lambda}^\dagger$ are annihilation and
creation operators for photons with wave vector $k$ and polarization 
$\lambda$, $\hat{S}_j^+$ and $\hat{S}_j^-=(\hat{S}_j^+)^\dagger $are raising 
and lowering operators for the internal levels in the usual 
spin $\frac{1}{2}$ representation, $\mathbf{d}_{eg}$ is the transition dipole 
moment from level g to level e, $\mathbf{e}_{k\lambda}$ are unit 
polarization vectors, and the allowed $k$ are integer multiples of $2\pi/L$
where $L$ is the length of a 1D microwave cavity resonant
with the hyperfine transition $g\leftrightarrow e$. Retaining coupling only to 
a single mode of wavelength
$\lambda=\frac{hc}{\epsilon_{eg}}$ resonant with the hyperfine transition, 
one obtains a Hamiltonian generalizing that of models I-III:
$\hat{H}=\hbar cq\hat{N}_q+\sum_{j=1}^N
[-\frac{\hbar^2}{2m}\frac{\partial^2}{\partial x_j^2}+v(x_j)+
\epsilon_{eg}\hat{S}_j^z
+i\gamma L^{-1/2}(\hat{S}_j^+\hat{b}_qe^{iqx_j}-\text{H.c.})]$
where $q=\frac{\epsilon_{eg}}{\hbar c}$, 
$\hat{N}_q=\hat{b}_q^{\dagger}\hat{b}_q$, 
$\gamma=\sqrt{\hbar cq}\mathbf{d}_{eg}\cdot\mathbf{e}_q$, a constant term 
$\frac{N}{2}(\hbar cq+\epsilon_g+\epsilon_e)$ has been dropped, and the
external potential is taken here to be that of a longitudinal harmonic trap,
$v(x)=\frac{1}{2}m\omega^2 x^2$.
To obtain the ground state with both coupling to the electromagnetic field
and TG and/or FTG interatomic interactions, one can first let the above 
Hamiltonian act on the space of ideal Fermi gas model states 
$\Psi_{\alpha M}^F$, find the ground state $\Psi_{0M}^F$ in that space, and 
then find the interacting ground state $\Psi_0^B$ by application of the 
mapping (\ref{1960map}). 

The length of typical cigar-shaped ultacold gas traps of high aspect ratio is 
many orders of magnitude smaller than relevant microwave wavelengths,  
so it is an excellent approximation to replace $e^{iqx_j}$ by unity
(dipole approximation for the whole $N$-atom system), leading to a zero-order 
Hamiltonian $\hat{H}_0=\hat{H}_{\text{trans}}+\hat{H}_{\text{Dicke}}$ where the
translational Hamiltonian is  
$\hat{H}_{\text{trans}}=\sum_{j=1}^N[-\frac{\hbar^2}{2m}
\frac{\partial^2}{\partial x_j^2}+v(x_j)]$ and the remainder is a resonant 
Dicke model,
$\hat{H}_{\text{Dicke}}=\epsilon_{eg}(\hat{N}_q+\hat{S}^z)
+i\gamma L^{-1/2}\sum_j(\hat{S}_j^+\hat{b}_q-\text{H.c.})]$ where 
$\hat{S}^z=\sum_j\hat{S}_j^z$ and the resonance condition 
$\hbar cq=\epsilon_{eg}$ has been inserted. There is strong 
statistical field-atom coupling since the eigenstates of 
$\hat{H}_{\text{trans}}$ are those of an ideal Fermi gas which must be 
antisymmetric under combined space-spin exchange
$(x_j,s_j)\leftrightarrow (x_\ell,s_\ell)$.

The ground state of $\hat{H}_{\text{Dicke}}$
in the thermodynamic limit is the direct product of a field-independent
state and a Glauber coherent state for the electromagnetic field 
\cite{WanHio73,HepLie73b}. This amounts to replacement of
the photon annihilation and creation operators by c-numbers $\beta_q$ and
$\beta_q^*$. The ground state energy is invariant under a gauge transformation
of the first kind $\beta_q\to\beta_qe^{i\theta}$ with $\theta$ arbitrary;
a convenient choice here is $\beta_q=-i\sqrt{N}f_q$ where $f_q$ is real, 
nonnegative, and independent of $N$. The corresponding reduced
Hamiltonian is $\hat{H}_0=\epsilon_{eg}Nf_q^2+\sum_{j=1}^N
[-\frac{\hbar^2}{2m}\frac{\partial^2}{\partial x_j^2}+v(x_j)+
\epsilon_{eg}\hat{S}_j^z
+\gamma\sqrt{\rho_q}f_q(\hat{S}_j^{+}+\hat{S}_j^{-})]$ where $\rho_q=N/L$. The 
single-atom energies $\epsilon_n^\pm$ and eigenstates $\phi_n^\pm$ can be found
by a $2\times 2$ matrix diagonalization for each j, with the results 
$\epsilon_n^\pm=n\hbar\omega\pm\frac{1}{2}
\sqrt{\epsilon_{eg}^2+4\gamma^2\rho_qf_q}$ and 
$\phi_n^\pm(x,s)=u_n(x)w_n^\pm(s)$ where $n=j-1=0,1,2,\cdots$, $u_n(x)$ are the
harmonic oscillator eigenfunctions, and 
$w_n^\pm(s=\frac{1}{2})/w_n^\pm(s=-\frac{1}{2})=(\epsilon_{eg}
\pm\sqrt{\epsilon_{eg}^2+4\gamma^2\rho_qf_q})/2\gamma\sqrt{\rho_qf_q}$. 
The quantum phase transition previously described in the 
presence of an external magnetic field $B$ will still occur, but now
at a higher value of $B$ since $\epsilon_n^+-\epsilon_n^->\epsilon_{eg}$.

The above assumes that the trapped ultracold gas is bathed in a 
constant microwave field which is supplied and controlled externally. If 
it is instead contained in a perfectly reflecting cavity 
with no external microwave source, one can investigate the possibility of
a Dicke thermal phase transition to a state with a self-generated
superradiant microwave field. Minimizing the grand canonical free energy 
$-\beta^{-1}\ln\text{Tr}\ e^{-\beta(\hat{H}_0-\mu\hat{N})}$ of the ideal
Fermi gas of models I-III with respect to $f_q$ \cite{WanHio73,HepLie73b} 
yields the following condition for determining $f_q$:
$2N\epsilon_{eg}f_q=\frac{\gamma^2\rho_q}{\sqrt{\epsilon_{eg}^2
+4\gamma^2\rho_qf_q}}\sum_{n=0}^\infty(g_n^--g_n^+)$ where
$g_n^\pm=[1+e^{\beta(\epsilon_n^\pm-\mu)}]^{-1}$ and $\mu=\epsilon_F$
is determined from $\sum_{n=0}^\infty(g_n^++g_n^-)=N$. Even for the much
simpler case of fixed atoms treated classically (no Fermi sea), solution
for the superradiant phase transition is nontrivial, so I shall not proceed 
further now. However, by comparison 
with that case \cite{WanHio73,HepLie73b} it is reasonable to conjecture that 
there is no phase transition for small values of the ratio
$\gamma\sqrt{\rho_q}/\epsilon_{eg}$ but a thermal transition to a 
superradiant phase 
for large values of this ratio. Recalling that $\epsilon_{eg}$
can be made arbitrarily small by tuning an external magnetic field $B$, 
it seems likely that this model will exhibit a superradiant phase transition 
for sufficiently large $B$, which may be coupled to the previously described
cooperative rearrangement of the translational ground state.
Note that arguments against existence of a superradiant transition in real
atomic systems \cite{Rza75Bia79} do not apply here, since $\epsilon_{eg}$ 
includes a negative Zeeman shift due to $B$.
\begin{acknowledgments}
I am grateful to Gonzalo Muga for detailed discussions of \cite{MCLM07}, 
to Ewan Wright for helpful suggestions, 
and to the U.S. Office of Naval Research for partial support of my research. 
\end{acknowledgments}
\end{document}